\documentstyle[12pt,epsf,epsfig]{article}
\def\ffrac#1#2{\textstyle{#1\over#2}\displaystyle}

\begin{document} 

\title{Conformal Invariance in Percolation, Self-Avoiding Walks, and
Related Problems\footnote{
Plenary talk given at the International Conference on Theoretical
Physics, Paris, July 2002.}}
\author{John Cardy\\
Theoretical Physics\\
1 Keble Road, Oxford OX1 3NP, United Kingdom\\
\& All Souls College, Oxford\\}
\maketitle
\begin{abstract}
Over the years, problems like percolation and self-avoiding walks have
provided important testing grounds for our understanding of the nature
of the critical state. I describe some very recent ideas, as well as some
older ones, which cast light both on these problems themselves and
on the quantum field theories to which they correspond. These ideas
come from conformal field theory, Coulomb gas mappings, and stochastic Loewner
evolution.
\end{abstract}

\begin{figure}
\epsfxsize=9.25cm
\epsfbox{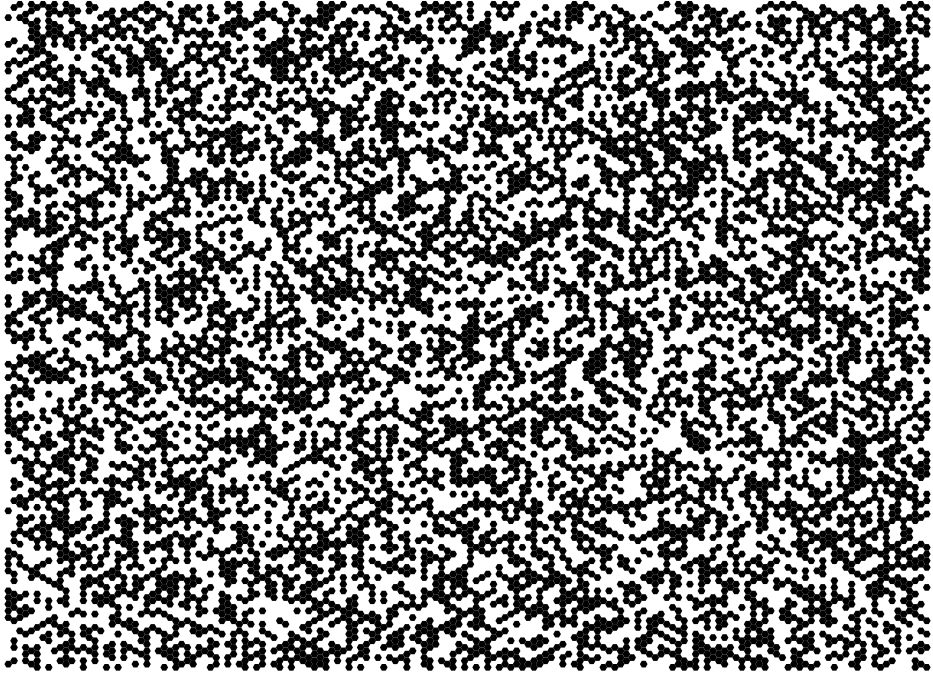}
\caption{Critical site percolation on the triangular lattice.
Each hexagon is independently coloured black or white with probability
$\frac12$. Is there a path on neighbouring black hexagons which connects
the left and right sides of the rectangle?}
\label{perc}
\end{figure}

\begin{figure}
\centerline{
\epsfxsize=12cm
\epsfbox{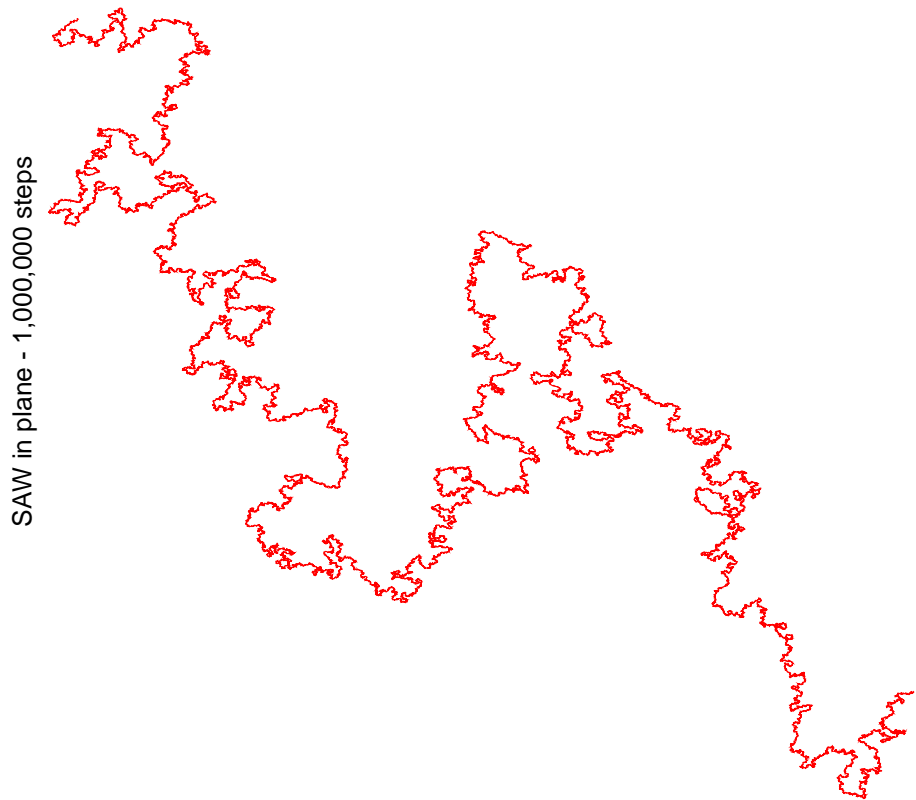}}
\caption{A typical self-avoiding walk. In this case the lattice is too
fine to be visible.}
\label{saw}
\end{figure}

This talk is about
`geometric' critical phenomena.
These are random spatial processes, where either (1) the probability
distribution is determined by equilibrium statistical mechanics,
and we ask questions about geometrical properties, 
or (2) the probability distribution is itself geometrical in nature.
The simplest example of (1) is clustering in
percolation (see Fig.~\ref{perc}), in which the probability
distribution is trivial (in this case sites of a triangular lattice
are independently coloured black or white with equal probability,) but
we ask questions like whether there exists a path on (say) the black
sites connecting opposite edges of a large rectangle.
The paradigm example of (2) is the ensemble 
self-avoiding walks (SAWs) of a fixed (large) length, all weighted equally 
(see Fig.~\ref{saw}). In this case one might ask, for example,
questions about the distribution of the distance between the ends.
Other examples abound: for example the clusters formed by the spins in a
critical Ising model, or the boundaries of the Fortuin-Kasteleyn clusters
in the Potts model.

Note that although percolation is relevant to disordered media 
and SAWs relevant but to polymer physics, the
emphasis in this talk
is on understanding the nature of their fractal
geometry. The outline is as follows:
first, we shall discuss the various 
expectations for `geometric' critical behaviour from conventional
critical behaviour, largely based on well-known mappings between these
types of problem. Since, in two dimensions, critical behaviour is
described by conformal field theory, we may deduce all sorts of
(non-rigorous) results, and we shall describe some of these.
A related set of non-rigorous techniques is based on a mapping of
cluster boundaries to a height models and then to a Coulomb gas, and
we shall mention these.

But the main point of my talk will be to bring to the attention of 
theoretical physicists relatively new ideas which, it so happens, were
developed by mathematicians, based on the
direct construction of continuum limit of cluster boundaries, and known
as SLE. I will describe how these provide
rigorous and new results for percolation, as an example. I finish with
some (slightly) provocative conclusions about the usefulness of rigorous
methods and of the traditional approach to quantum field theory, in the
study of critical phenomena in general.
\subsection*{`Geometric' {\it vs.}  `conventional' critical behaviour}

Much of our intuition about geometric critical behaviour is based on two
well-known mappings:
\begin{enumerate}
\item (Fortuin-Kasteleyn\cite{FK})  $Q$-state Potts 
model $\Leftrightarrow$\\
Random Cluster Model: the Potts model is a generalisation of the Ising
model in which `spins' at the sites of a lattice can each take one of 
$Q$ values. Initially, $Q$ must be an integer $\geq2$, but the partition
function may also be written
\begin{displaymath}
Z=\langle Q^{|{\rm clusters}|}\rangle_{\rm percolation}
\end{displaymath}
where the clusters are weighted as in percolation, with the parameter
$p$ being simply related to the temperature in the Potts model.
Evidently, the limit $Q\to1$  reproduces percolation.
Although in this limit the partition function (with free boundary
conditions) is trivially equal to 1, the correlations are nontrivial.
\item (de Gennes\cite{dG})
O$(n)$ model $\Leftrightarrow$ Self-Avoiding Loops:
This model similarly generalises the Ising model to $n$-component
spins and a hamiltonian which is invariant under O$(n)$ rotations.
The partition function may be expressed as that of a gas of
non-intersection loops:
\begin{displaymath}
Z=\langle n^{|{\rm loops}|}\rangle_{\rm loop\ gas}
\end{displaymath}
Evidently the case
$n=1$ corresponds to the Ising model, while the limit
$n\to0$ leaves a single self-avoiding loop.
In fact the O$(n)$ model has two types of critical behaviour for $n<2$: one
corresponding to the dilute phase, and a \em dense \em critical phase.
In the dense phase, the loops are in the same universality class as the
hulls of F-K clusters with $Q=n^2$. 
\end{enumerate}
From these correspondences, we can build a dictionary which relates
geometrical properties to more conventional thermodynamic quantities.
For example
\begin{eqnarray*}
{\rm Cluster\ size}&\Leftrightarrow&{\rm
susceptibility}\propto(p-p_c)^{-\gamma(Q)}\\
{\rm Radius}&\Leftrightarrow&\mbox{correlation length}
\sim{\rm mass}^{\nu(n)}
\end{eqnarray*}
\subsection*{Critical Behaviour \& Euclidean Field Theory}
It has been realised since the late 1960s that the scaling limit of 
an isotropic system near a continuous phase transition is a euclidean
quantum field theory.
If we take a near-critical lattice model, such that the 
correlation length $\xi\gg$ the lattice spacing
$a$, the following limit exists:
\begin{displaymath}
\langle\phi(r_1)\ldots\phi(r_N)\rangle_{\rm QFT}
= \lim_{a\to0,\ \xi\ {\rm fixed}}a^{-nx_\phi}
\langle S(r_1)\ldots S(r_N)\rangle_{\rm lattice}
\end{displaymath}
where $\phi(r)$ is a local quantum field, and $S(r)$ is the
corresponding lattice quantity. The non-trivial power $x_\phi$ is the
scaling dimension of $\phi$.
This correspondence is rooted in an 
emphasis on correlation functions of \em local \em (or
quasi-local) operators and their algebra encoded in the 
operator product expansion (and is therefore not always the best tool
to investigate other quantities.)
It has been \em proved \em in very few examples, but if \em assumed \em
it has many powerful consequences: the Renormalisation Group, universality,
and, in particular \em scaling\em:a the property that the critical
exponents describing
off-critical behaviour of thermodynamic quantities are simply related
to those describing
decay of correlation functions at the critical point. This means that,
for many purposes, we may restrict ourselves to studying the behaviour
at the critical point. This means that the corresponding quantum field
theory is massless: a \em conformal field theory\em.
\subsubsection*{Conformal Field Theory}
Conformal field theory provides a very powerful tool to study critical
behaviour, especially in two dimensions:
\begin{itemize}
\item in local \em classical \em field theories, scale invariance
implies that the trace of the stress tensor
$T^\mu_\mu$, vanishes, and this by itself implies conformal invariance. 
\item CFT assumes this holds (up to conformal anomaly $c$) in
the full theory including fluctuation effects.
\item In CFT, unlike normal QFT, there is a 1-1 correspondence between
local operators and states of the Hilbert space, and the spectrum of
these is usually discrete rather than continuous.
\item For $d=2$, these transform according to irreducible
representations of an
infinite-dimensional Virasoro algebra. Classifying these essentially
classifies all possible CFTs, that is, all universality classes. 
\item There is, however, a problem: given some critical lattice model, to
which CFT does it correspond?
\item this was answered in part in the work of
Friedan, Qiu and Shenker\cite{FQS}, who showed that in theories having 
reflection positivity (eg. $Q=2,3,4$ Potts models, or $n=1,2$ in
the O$(n)$ model), we should look for 
\em unitary \em representations, and at least when $c<1$ 
this leads to a discrete series of possibilities. Moreover, the necessary
decoupling of null states in these theories leads to
linear differential equations for correlation functions
(Belavin, Polyakov, Zamolodchikov\cite{BPZ}.)
\end{itemize}

But percolation, SAWs, and related models are not unitary: in fact they
have partition function $Z=1$  ($c=0$) even though 
their correlations are nontrivial.
In fact non-unitary $c=0$ CFTs  are very poorly understood. 
Nevertheless they are important not just for
percolation and SAWs, but for all critical problems with quenched
disorder (e.g. the quantum Hall plateau transition, a major unsolved
problem.)

\subsection*{The Crossing Formula in Percolation}
A recurring theme in this talk will be the following problem
(see Fig.~\ref{crossing}):
\begin{figure}
$$
\epsfxsize=10cm
\epsfbox{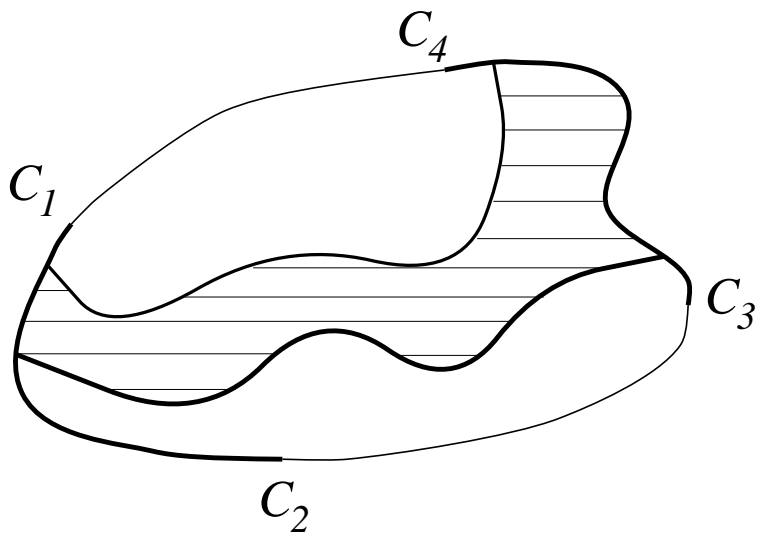}
$$
\caption{Schematic represensation of a crossing cluster which connects
the arcs $C_1C_2$ and $C_3C_4$. Of course in reality this is a fractal
object with a fractal boundary.}
\label{crossing}
\end{figure}

$\bullet$ Given a simply connected region $D$ of the plane, with
suitably smooth boundary $\partial D$ with 4 marked points $C_j$, what
is the probability of a spanning cluster connecting $C_1C_2$ with
$C_3C_4$ (in the limit lattice spacing $\to0$)? 

The following conjecture was made (Cardy\cite{JC}), based on ideas of
CFT and the mapping to the $Q\to1$ limit of the Potts model: the
Riemann mapping theorem allows us to conformally 
map interior of $D$ into the unit disc, with the
marked points $C_j\to z_j$. Then the 
crossing probability depends only the anharmonic
ratio $\eta=z_{12}z_{34}/z_{13}z_{24}$ and is 
\begin{displaymath}
{\Gamma(\ffrac23)\over\Gamma(\ffrac43)\Gamma(\ffrac13)}\,
\eta^{\ffrac13}{}_2F_1(\ffrac13,\ffrac23,\ffrac43;\eta)
\end{displaymath}
The argument depends on:
\begin{itemize}
\item assuming that the
scaling continuum limit exists and is given by a CFT with
$c=0$;
\item realising that crossing probability is related to a difference of
partition functions of the $Q\to1$ Potts model, with different boundary
conditions (Fig.~\ref{zz});
\begin{figure}
\begin{center}
\epsfxsize=10cm
\epsfbox{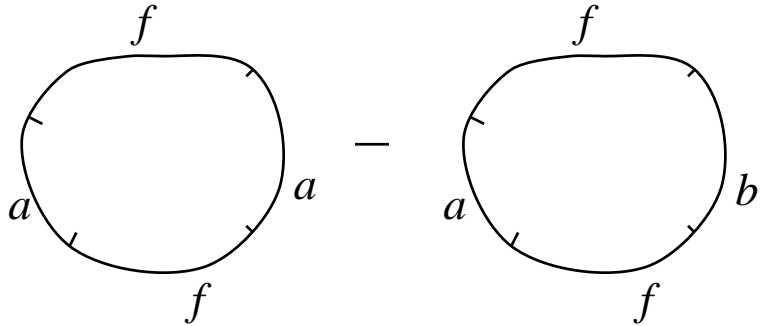}
\end{center}
\caption{The crossing probability is a difference of Potts model
partition functions, in which the Potts spins are fixed into the same
($a$), or different ($a$ and $b$), states along the two arcs $C_1C_2$
and $C_3C_4$.}
\label{zz}
\end{figure}
\item realising that states of the CFT induced by changes in the
boundary condition
(`boundary condition changing operators') also should correspond to
Virasoro representations;
\item guessing the right representation, and hence deducing the
appropriate differential equation.
\end{itemize}

The formula has been numerically verified to high precision, 
but it is hard to see how to make arguments rigorous, or
to go beyond them.
\subsection*{Cluster Boundary Approach}

Instead of thinking about clusters, in the Potts language, it is
sometimes easier to think about cluster boundaries, or hulls, using
the O$(n)$ language. There are really two different approaches here:
the older Coulomb gas arguments, and the more recent ideas of SLE.

\subsubsection*{`Coulomb gas' method}
(den Nijs\cite{dN}, Nienhuis\cite{Nien}, Duplantier \& Saleur\cite{DupSal}, 
Kondev\cite{Kond}, $\ldots$)
The elements of these arguments are:
\begin{itemize}
\item thinking of cluster boundaries in random cluster model (or loops in O$(n)$
model) as a gas of (unoriented) closed loops. 
\item randomly orientate each loop: each configuration of oriented loop
than maps onto one of a \em height model\em, with
degrees of freedom $h(r)\in$ integers, on the dual lattice (see
Fig.~\ref{loops}):
\begin{figure}
\begin{center}
\epsfxsize=12cm
\epsfbox{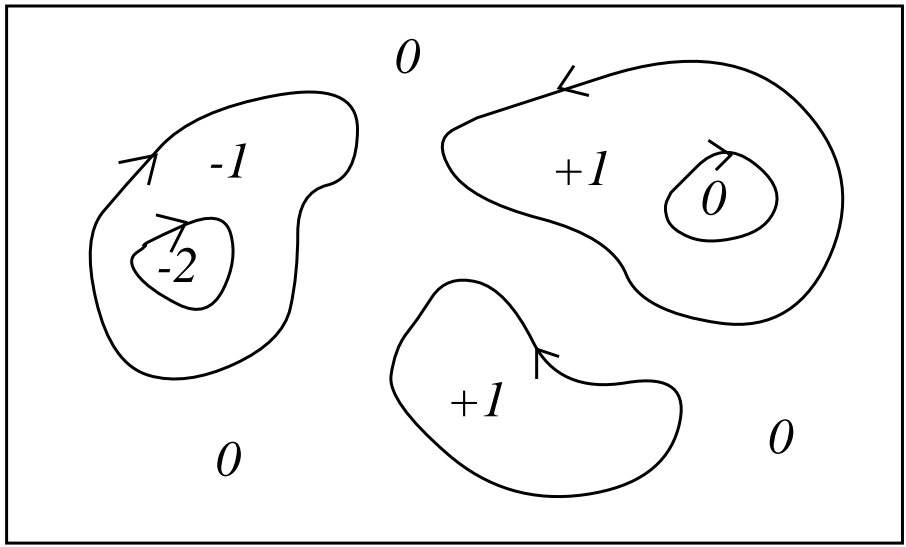}
\end{center}
\caption{Oriented loops and the mapping to configurations 
of the height model.}
\label{loops}
\end{figure}
\item the factors of $Q$ (resp.~$n$) can be associated with local (but 
in general complex!) Boltzmann weights;
\item \em assume \em in the continuum limit that $h(r)$ takes values
in the real numbers, and that the
measure converges to a gaussian
$\exp(-g{\displaystyle\int}
(\partial h)^2d^2\!r)$;
\item this gives a $c=1$ CFT, with, however, various bells and whistles
like a charge at $\infty$, screening charges,
etc., which make it nontrivial;
\item within this formulation critical exponents are 
calculable, eg $\nu_{\rm
SAW}=\frac34$ and
$\nu_{\rm perc}=\frac43$;
\item \em but \em correlation functions are ambiguous, and, more
seriously, it has proven very hard to make this approach
rigorous.
\item nevertheless new results are still emerging from this method: 
eg distribution of \em internal
areas \em of loops: density $n(A)$ of large loops with area $>A$
\begin{displaymath}
n(A)\sim C/A\qquad\mbox{$C$ universal}
\end{displaymath}
where\cite{CZ}
\begin{eqnarray*}
C_{\rm perc}=1/8\surd3\pi&=&0.0229720\qquad\mbox{predicted}\\
              &=&0.022972(1)\quad\mbox{measured}
\end{eqnarray*}
\end{itemize}
\subsubsection*{Dynamical description of cluster boundaries (SLE)}
\begin{figure}
\centerline{
\epsfxsize=13cm
\epsfbox{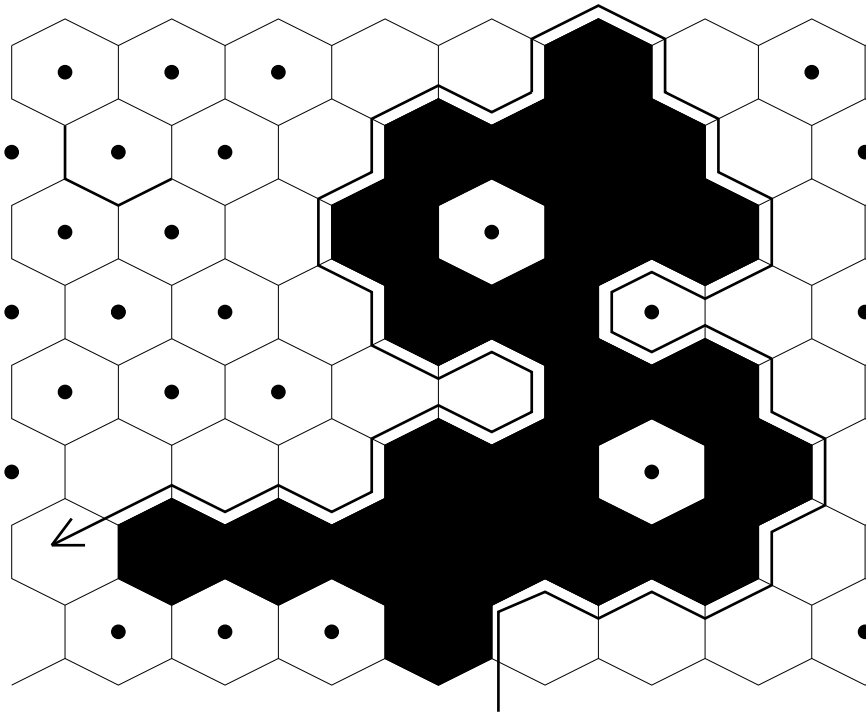}}
\caption{The process in which a random walker lays down the randomness
at it progresses.}
\label{walker}
\end{figure}
Although we usually think of percolation in terms of first laying down a
particular random configuration of black or white sites, and then
identifying the clusters and the boundaries of the clusters, it is
statistically  equivalent to construct these boundaries as certain
random walks, where the walker lays down the random configuration 
as it goes. More specifically, this `exploration' process involves
random walker laying down the
configuration as it moves: black sites to the left, white to the right 
(see Fig.~\ref{walker}).
Notice that the path automatically reflects from itself, and also from
the boundary if we choose the correct boundary conditions (black sites
to the left of the starting point, white to the right.) 

How should one characterise the continuum limit of these paths (assuming
this exists)? In what sense is it conformally invariant?
These were the questions addressed by
Schramm\cite{Schramm} and Lawler, Schramm and Werner\cite{LSW}, 
and the answer is:
\subsection*{Stochastic Loewner Evolution(SLE)}
For definiteness, consider the half-plane 
with black sites along $x<0$ and white sites along $x>0$. The walker
starts at the origin, as shown in Fig.~\ref{2f1}: 
\begin{figure}
$$
\epsfxsize=9.25cm
\epsfbox{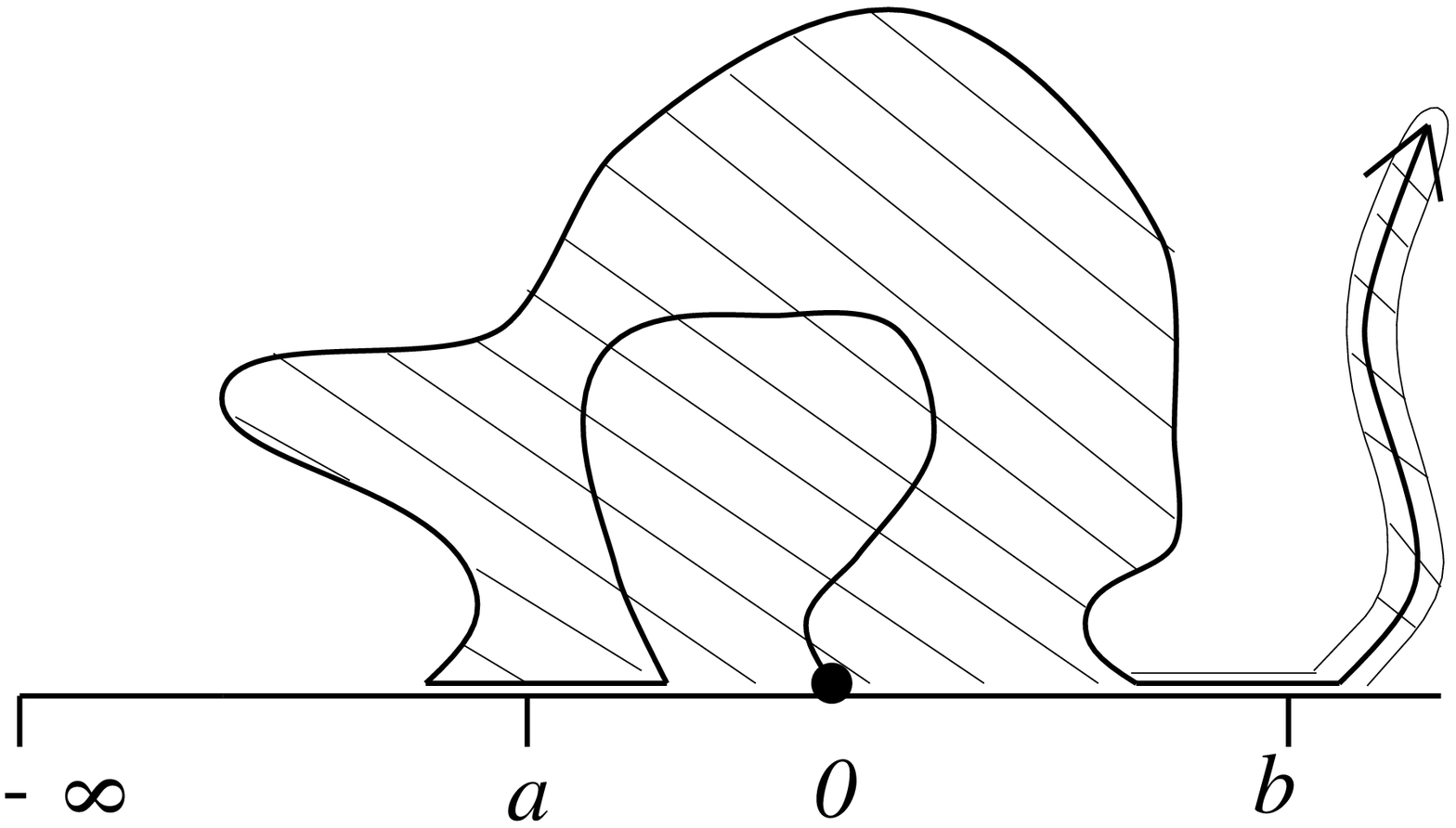}
$$
\caption{Schematic representation of SLE. At any given time, the path,
and the region between it and the real axis, form an excluded
part of the upper half plane. The complement of this is conformally 
mapped by the
function $g_t(z)$ into the whole upper half plane. This function obeys
the stochastic Loewner equation.}
\label{2f1}
\end{figure}

Rather than trying to write an equation for the path, consider the 
conformal mapping $z\to g(z;t)$
which sends \{region of the half-plane which has not
been excluded by the path\} $\to$ \{upper half-plane\}.
Then, instead of the dynamics of the path, we may think about
dynamics on conformal mappings. In particular, Schramm\cite{Schramm} conjectured
that the continuum limit of the percolation exploration process
corresponds to the Loewner equation 
\begin{displaymath}
{\partial g(z;t)\over\partial t}={2\over g(z;t)-a(t)}
\end{displaymath}
This has several important properties:
\begin{itemize}
\item if $a(t)$ is a real continuous function, the excluded region of the
half-plane grows with increasing $t$;
\item from requirements of scaling and locality, we conclude that
$a(t)$ must be Brownian motion, ie $\dot a=\zeta(t)$ with
$\overline{\zeta(t)\zeta(t')}=\kappa\delta(t-t')$.
(Rohde \& Schramm\cite{RS}) 
\begin{eqnarray*}
{\rm If}\quad0\leq\kappa\leq 4&&\mbox{path is simple}\\
4<\kappa<8&&\mbox{it touches itself}\\
8<\kappa&&\mbox{it fills space}
\end{eqnarray*}
\item the fractal dimension of the path is $=1+\kappa/8$ (for $\kappa<8$).
\item \em only \em for
$\kappa=6$ does the SLE path not `feel' where the boundary of
the domain is as long as it does not hit it, as one expects for
percolation with uncorrelated site probabilities. This led to the
conjecture:
\begin{center}
{\bf SLE$_6$ is the conformally invariant continuum limit of percolation
cluster boundaries}
\end{center}
\item one can actually \em compute \em with SLE: it involves arguments
quite familiar to theoretical physicists working in a different area,
namely stochastic processes. In this way, one obtains
all previously conjectured
critical exponents at $p_c$ (and with the help of rigorous scaling
relations (Kesten\cite{Kest}), exponents away from $p_c$),
including multifractal irrational but algebraic exponents (related to 2d
quantum gravity (Duplantier\cite{Dup}), and some new ones, eg the
\item \em backbone exponent \em 
(Lawler, Schramm, Werner\cite{Back}), the fractal 
dimension of the part of the
infinite cluster which would carry electric current. This is given by the
lowest eigenvalue of a 2d Dirichlet problem (and is probably not a
rational or even an algebraic number, which makes its derivation by
CFT or Coulomb gas methods a real challenge.) 
\item once SLE is assumed to describe the limit of percolation hulls,
one easily gets and the crossing formula. 
As illustrated in Fig.~\ref{2f1}, if we map the region into the upper
half plane so that the arcs $C_1C_2$ and $C_3C_4$ map to
$(-\infty,a)$ and $(0,b)$ respectively, then 
\begin{eqnarray*}
&&Pr({\rm white\ crossing\ } (-\infty,a)\leftrightarrow(0,b)\\
&&= Pr(\mbox{$a$ gets excluded before $b$})
\end{eqnarray*}
\item this gives the same ${}_2F_1$ formula as conjectured from CFT.
\end{itemize}
However, all this is so far dependent on Schramm's conjecture.
\subsubsection*{The missing link: Smirnov's proof of the crossing formula}
Smirnov\cite{Smir} proved that the crossing formula holds for the continuum limit
of site percolation on a triangular lattice, and thereby that
SLE$_6$ 
\em is \em the continuum limit of percolation cluster boundaries.
Therefore all the results derived from SLE$_6$ are \em rigorous\em.
\begin{itemize}
\item {} First, it was observed by Carleson that the crossing 
formula is simple in an equilateral triangle (see Fig.~\ref{triangle}).
\begin{figure}
\begin{center}
\epsfxsize=9.25cm
\epsfbox{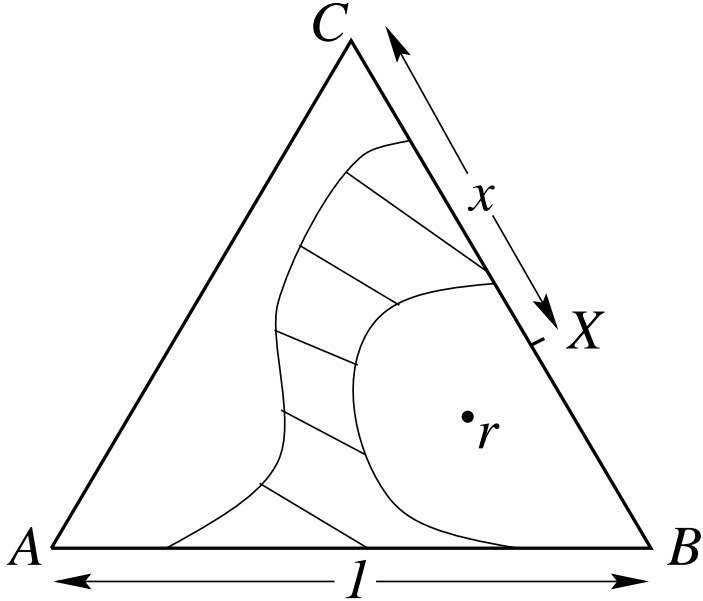}
\end{center}
\caption{Carleson's version of the crossing formula in an equilateral
triangle, and Smirnov's generalisation: what is the probability that
there is a cluster which connects $AB$ with $BC$, at the same time
disconnecting the point $z$ from $AC$?}
\label{triangle}
\end{figure}
\item Note that the formula as proposed 
is the boundary value of an analytic function $P(z)$: so what 
is its interpretation for $z$ not on the boundary?  Smirnov proposed
studying 
$P(z)= Pr(z$ separated from $C_1C_4$ by at least one cluster spanning from
$C_1C_2$ to $C_2C_4$). Then:
\item on the triangular lattice, 
$P(z)$ satisfies linear
relations, which mean that its continuum limit exists, and in fact just
is the real part
of a harmonic function, ie, a solution of Laplace's equation.
It is the rather unusual boundary conditions which then determine that
$P(z)\propto$ distance from $C_1C_4$, which leads directly
to the crossing formula when $z$ is taken to lie on the boundary..
\end{itemize}
\subsubsection*{Other values of $\kappa$}
It turns out that other values of $\kappa$ in SLE correspond to
different values of $n$ in the O$(n)$ model, for example
\begin{itemize}
\item Self-Avoiding Walks: if we use the principle that
uniform measure on set of simple
paths must remain uniform
when restricted to a subset, and assume the continuum limit is
SLE$_\kappa$, then we are led to
$\kappa=\frac83$. This reproduces all the conjectured
results for SAWs (and more) (Lawler, Schramm, Werner\cite{LSWsaw}). 
\item the critical Ising model is
conjectured to be described by SLE$_3$. This leads to all the
standard results and some new ones: eg, if we take an Ising model inside
a simply connected
region of the plane, with boundary conditions that the spins are up on
one segment of the boundary, and down on its complement (so that there
exists a domain wall crossing the region, see Fig.~\ref{ising}), what is
the probability that the domain passes above a given point $z$?
The result is conformally invariant, and an explicit formula is
provided by SLE (Schramm\cite{Schr}). Of course these results still need the
analogue of Smirnov's proof to make them completely rigorous, but they
appear to be completely beyond the reach of traditional approaches to
the Ising model, which focus on spin correlation functions.
\begin{figure}
\begin{center}
\epsfxsize=8cm
\epsfbox{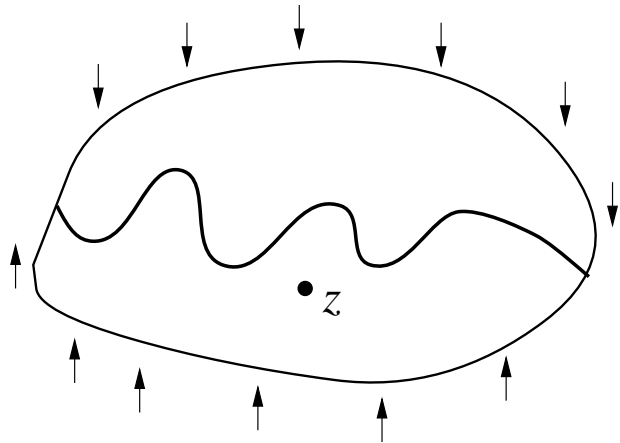}
\end{center}
\caption{The Ising model in a domain with $+$ boundary conditions on 
one segment of the boundary, and $-$ boundary conditions on the
complement: what is the probability that the consequent domain wall
passes above the point $z$?}
\label{ising}
\end{figure}
\end{itemize}
\subsection*{Final Remarks}
I shall finish with a couple of possibly provocative remarks. The first
concerns progress in rigorous results for critical behaviour.
The first such nontrivial results were of course obtained for the Ising
model in 1944 by Onsager, followed by the work of Yang, and this gave
rise to a whole industry of computations of correlation functions and
other quantities in this and variant models. The next step was perhaps
Lieb's 1967 solution of 6-vertex type models, then Baxter's 
solution of the 8-vertex model in 1971. The methods used by Baxter
and others (Bethe ansatz, commuting transfer matrices, corner transfer
matrices) are powerful and elegant, and they produce `exact' results,
but they are not, in general, fully rigorous. 

Now we finally have, thanks to the work of 
Kesten, Smirnov, Lawler, Schramm, Werner, and others
\em rigorous \em results for a  different sort of two-dimensional critical 
behaviour, namely percolation. The methods used are completely
different, relying on identifying the continuum limit explicitly, and
performing computations directly in that limit, rather than taking
the limit of lattice results. As a result, only universal quantities are
calculated. The methods focus on the geometrical aspects of the measure,
rather than the correlation functions of local operators. It is quite
likely that in the next few years we shall see similar rigorous results
for a whole set of such models, at least in two dimensions.

Of course, particularly at a meeting on Theoretical Physics, one might
question the necessity of mathematical rigour. Are we not only
interested in the answers, and in the physical picture which they
convey? Of course this is true, but the historical fact remains
that theoretical physicists in this subject have been perhaps too happy for
many years with their heuristic and incomplete arguments, which had
once been very fruitful but perhaps had reached a dead end. It took
mathematicians to question the true content of these arguments, to find
them wanting, and then to develop a completely independent approach, that
of SLE. The result is that we now are beginning to have a new physical way of
understanding the origins of conformal invariance and of the emergence
of non-trivial critical exponents, quite separate from that of the
renormalisation group. The methods of SLE could have been developed by
physicists -- after all they are based on Brownian motion -- but the
fact is that they were not. This affords a strong example for
continuing to make theoretical physics as mathematically rigorous as
possible.

The second point concerns 
progress in quantum field theory, especially from the point of
view of critical behaviour. Historically, the development of the 
renormalisation group point of view in 1969 by Wilson and others
gave answers to many outstanding questions at the time. 
It gave a non-rigorous framework in which to understand many important
features of critical phenomena, such as universality and scaling,
and to do approximate computations.
From 1984 the methods of conformal field theory, originally developed
for string theory, were brought in, and they provided 
a plethora of exact but non-rigorous results in two dimensions.
But even today field theory is still tied to particle physics
ideas. We see that both in the way it is taught and in the continuing
focus on the objects of interest to particle theory:
namely correlation functions of local fields. While this has been
successful in the past, it is
ill-suited to study other objects (eg the crossing formula, which
required the introduction of `boundary-condition changing' operators to
be expressed in the old language.) One might ask whether
QFT has outlived its usefulness to generate new results, or whether
(rather like string theory)
it can reinvent itself in the 21st century, in a different and more
powerful form -- perhaps
\vspace{1cm}
\begin{center}
{\bf QFT as fractal geometry?}
\end{center}
\vspace{1cm}
\noindent {\it Note added.} Very recently Bauer and Bernard\cite{BB}
have, among other things, exposed the relationship between Virasoro null
vectors and SLE, thus establishing the connection with the CFT approach
to the crossing formula.

This work was supported in part by EPSRC Grant GR/J78327. I thank
R.~Ziff for help in producing Fig.~\ref{perc}, and
T.~Kennedy for permission to reproduce Fig.~\ref{saw}.
\end{document}